\documentclass[aps,prc,showpacs,twocolumn,superscriptaddress]{revtex4-1}

\usepackage{lipsum}
\usepackage{color}
\usepackage{amsmath,amsfonts,amssymb,amsthm}

\usepackage{graphicx}
\graphicspath{{fig/}}
\usepackage{siunitx}
\usepackage{hyperref}
\hypersetup{hidelinks,colorlinks=false,breaklinks=true,urlcolor=ocre}

\usepackage{dcolumn}
\usepackage{bm}

\begin{document}

\preprint{APS/123-QED}

\title{The impact of neutral impurity concentration on charge drift mobility in p-type germanium}

\author{H. Mei}
 \affiliation{Department of Physics, The University of South Dakota, Vermillion, South Dakota 57069, USA}
\author{D.-M. Mei} 
 \email{Corresponding author.\\Email:Dongming.Mei@usd.edu}
 \affiliation{Department of Physics, The University of South Dakota, Vermillion, South Dakota 57069, USA}%
\affiliation{School of Physics and Optoelectronic, Yangtze University, Jingzhou 434023, China}
\author{G. Wang}
\affiliation{Department of Physics, The University of South Dakota, Vermillion, South Dakota 57069, USA}
\author{G. Yang}
\affiliation{Department of Physics, The University of South Dakota, Vermillion, South Dakota 57069, USA}





\begin{abstract}
We reported a new result of the neutral impurity scattering of holes 
that has impact on the charge drift mobility in high purity p-type germanium crystals at 77 Kelvin. The charge carrier concentration, mobility and resistivity are measured by Hall Effect system at 77 Kelvin. We investigated the contribution to the total charge drift mobility from ionized impurity scattering, lattice scattering, and neutral impurity scattering with the best theoretical models and experimental data. Several samples with measured Hall mobility from the grown crystals are used for this investigation. With the measured Hall mobility and ionized impurity concentration as well as the theoretical models, we calculated the neutral impurity concentration by the Matthiessen's rule. As a result, the distributions of the neutral impurity concentrations with respect to the radius of the crystals are obtained. Consequently, we demonstrated that neutral impurity scattering is a significant contribution to the charge drift mobility, which has dependence on the concentration of neutral impurities in a given germanium crystal.
\end{abstract}

\maketitle

\section{\label{sec:intro}Introduction}
The charge drift mobility, depending on electric field, temperature, lattice and field orientation, is a critical parameter in the understanding of the rise time of charge pulses, which are often used to determine the sequence of gamma interactions inside a germanium detector, and to find the origin of gamma tracks for Gamma-Tracking experiments such as AGATA~\cite{agata} and GRETINA~\cite{gretina}. In the germanium-based neutrinoless double-beta decay experiments, such as GERDA~\cite{gerda} and Majorana~\cite{majorana}, the charge drift mobility plays a critical role in measuring the charge pulse shape~\cite{cooper, david, martin} that distinguishes background events (the multiple-site events from Compton scattering) from signal events (the single-site events from two electrons of neutrinoless double-beta decay process). The digital pulse shape is determined using the rise time of the charge pulse induced by the energy deposition in a given germanium detector. The rise time is proportional to the charge drift velocity, $\nu=\mu E$, where $\mu$ is the charge drift mobility and $E$ is electric field. To ensure that the detector performance of the digital pulse shapes is well understood, it is essential to simulate the digital pulse shapes to compare with the measured ones. In the Monte Carlo simulation, the electric field distribution can be accurately calculated using Poisson equation for an applied voltage and a given geometry of the detector, while the charge drift mobility is often treated as a constant. This can be true only if the charge drift mobility at 77 Kelvin is solely constrained by the lattice scattering, which is a constant at low fields, and for the $<$100$>$ axis for a given temperature. However, the charge drift mobility is usually governed by several scattering mechanisms including ionized impurity scattering, neutral impurity scattering, lattice scattering, and others. It is well known that the charge drift mobility can be impacted by temperature, crystal orientation, impurity concentration, defect concentration, and also electron and hole concentration~\cite{cmnj}. Among them, the scattering mechanisms play a very important role in determining the total charge drift mobility for a given crystal orientation. Traditionally, the lattice scattering and the ionized impurity scattering are used to determine the total charge drift mobility in the Monte Carlo simulation~\cite{cooper, david}. The neutral impurity scattering has not been considered to be an important contribution to the total charge drift mobility at 77 Kelvin and has been reported in some literature as having no direct impact on the electrical properties of germanium  crystals~\cite{dmbr}. In 1994, K. M. Itoh et al. found that the neutral impurity scattering is the dominant component of the charge drift mobility when the temperature is below 20 Kelvin~\cite{km}. Similar to other electrically active impurities, the neutral impurities can also act as scattering centers to impede the drift and diffusion of charge carriers under an electric field. As a result, the charge drift mobility is expected to decrease as the neutral impurities concentration increases~\cite{ce, tcm}. To demonstrate the variation of the charge drift mobility as a function of the concentration of neutral impurities, one must separate the portion of the charge drift mobility contributed by the ionized impurity scattering, lattice scattering, and other scatterings. This requires a comprehensive study of the contributions to the charge drift mobility due to the ionized impurity scattering, lattice scattering, and neutral impurity scattering, and others.
 In this work, the calculation of the charge drift mobility due to different scattering processes, and the impact of neutral impurity concentrations on the charge drift mobility are presented in section~\ref{s:mobi}, followed by the experimental results in section~\ref{s:exp}. Finally, we summarize our conclusions in section~\ref{s:conc}.

\section{Calculation of Charge Drift Mobility}
\label{s:mobi}
The charge drift mobility of germanium detectors depends on the rate at which charge carriers are scattered by impurity atoms and defects in the crystalline structure. The rate is the reciprocal of the relaxation time, which is the average time between collisions, and will be discussed later in terms of each scattering mechanism. The relationship between the relaxation time and the charge drift mobility is given by~\cite{brn}:
\begin{equation}
{\mu}=\frac{q\tau}{m^{\ast}},
  \label{mu}
\end{equation}
where $\mu$ is the charge drift mobility, $\tau$ is the relaxation time, $q$ is the elementary charge and $m^{\ast}$ is the effective mass of the charge carrier.

The total charge drift mobility of a charge carrier has contributions from at least five main scattering processes in a germanium detector: ionized impurity scattering, neutral impurity scattering, acoustic phonon scattering, optical phonon scattering and dislocation scattering~\cite{se}. Note that acoustical phonon scattering and optical phonon scattering belong to lattice scattering. 
The total charge drift mobility is impacted independently by the scattering mechanisms mentioned above and can be determined by using Matthiessen's rule~\cite{da}:
\begin{equation}
\frac{1}{\mu_{T}} = \frac{1}{\mu_{I}}+\frac{1}{\mu_{N}}+\frac{1}{\mu_{A}}+\frac{1}{\mu_{O}}+\frac{1}{\mu_{D}},
  \label{e:mu}
\end{equation}
where $\mu_{T}$ is the total charge drift mobility, $\mu_{I}$, $\mu_{N}$, $\mu_{A}$, $\mu_{O}$ and $\mu_{D}$ are the mobilities contributed by the scattering of ionized impurities, neutral impurities, acoustic phonons, optical phonons and dislocation, respectively. For p-type germanium, the majority charge carriers are holes. Since the existence of heavy and light holes, we have to consider the behaviors for each scattering mechanism by heavy and light holes separately. The effective mass $m^{\ast}$ is taken as $m_{h}^{\ast }$= 0.28 for heavy holes and $m_{ l}^{\ast }$= 0.044~\cite{sinj} for light holes, respectively. The ratio of heavy holes and light holes concentration $ {p_{h}}/{p_{l}}$ is given by $({m_{h}}/{m_{l}})^{3/2}$=16.05~\cite{dmbr}. Then by weighing the mobility of heavy holes $\mu_{h}$ and light holes $\mu_{l}$ according to their relative populations, we were able to calculate the individual total hole mobility for each scattering mechanism using the equation:
\begin{equation}
{\mu}=({\mu_{l}}{p_{l}}+{\mu_{h}}{p_{h}})/({p_{l}}+{p_{h}})=({\mu_{l}}+16.05{\mu_{h}})/17.05
 \label{e:com}
\end{equation}
 then the individual contributions are combined using eq.~\ref{e:mu}.

As indicated by eq.~\ref{e:mu}, the mobility due to neutral impurity scattering, $\mu_{N}$, can be evaluated if $\mu_{T}$, $\mu_{I}$, $\mu_{A}$, $\mu_{O}$ and $\mu_{D}$ are known. Furthermore, the neutral impurity concentration, $N_{n}$, can be estimated if the relationship between $N_{n}$ and $\mu_{N}$ can be established. 

\subsection{Ionized impurity scattering}
Ionized impurity scattering occurs when a charge carrier deviates from its trajectory by a Coulomb interaction as it gets close to an ionized impurity atom~\cite{dan}. The relaxation time was found by Conwell and Weisskopf using the Rutherford scattering formula~\cite{ec}:
\begin{equation}
\frac{1}{\tau }=\frac{2\pi N_{i}q^{4}}{(\varepsilon_\text{r}\varepsilon_0)^{2}m^{\ast 2}\nu ^{3}}ln(1+\frac{(\varepsilon_\text{r}\varepsilon_0)^{2}m^{\ast 2}\nu ^{4}}{4q^{4}N_{i}^{2}}),
\end{equation}
where $N_{i}$ is the ionized impurity concentration, $\varepsilon_{r}$=16 is the relative permittivity of germanium, $\varepsilon_0$ is the free-space permittivity, $m^{\ast}$ is the effective mass of the drifting charges in germanium, and $\nu$ is the charge carrier velocity. Then the mobility due to ionized impurity scattering $\mu_{I}$, can be calculated by the CW model~\cite{bh}:
\begin{equation}
\mu_{I}=\frac{128\sqrt{2}\pi^{1/2}(\varepsilon_\text{r}\varepsilon_0)^{2}(k_BT)^{3/2}}{m^{\ast 1/2}N_{i}Z^2q^3}/ln(1+\frac{144{\pi^{2}}(\varepsilon_\text{r}\varepsilon_0)^{2}k_B^{2}T^{2}}{Z^{2}q^{4}N_{i}^{2/3}}),
\label{e:mui0}    
\end{equation}

where $k_B$ is the Boltzmann constant, $T$ is the temperature in Kelvin and $Z=1$ is the effective charge number in germanium. A more accurate model was developed by Brooks and Herring since this CW model cuts off the small angle scattering~\cite{ns}, and does not take the effect on the potential of the distribution of space charges around the impurity atoms into account. In the Brooks-Herring theory~\cite{bh} or BH model, $\mu_{I}$, can be expressed as:
\begin{equation}
\mu_{I}=\frac{128\sqrt{2}\pi^{1/2}(\varepsilon_\text{r}\varepsilon_0)^{2}(k_BT)^{3/2}}{m^{\ast 1/2}N_{i}Z^2q^3}/ln\frac{24m^{\ast}\varepsilon_\text{r}\varepsilon_0(k_BT)^2}{N_{i}q^2\hbar^2}.
\label{e:mui1}
\end{equation}

If we replace all constants in eq.~\ref{e:mui1} with their corresponding values, eq.~\ref{e:mui1} is simplified as follows when $T$ = 77 Kelvin:
\begin{equation}
\mu_{I}=\frac{5.70\times 10^{20}}{m^{\ast 1/2}\cdot N_{i}}\left [ ln\frac{1.23\times 10^{19}\cdot m^{\ast } }{N_{i}} \right ]^{-1},
\end{equation} 
where $\mu_{I}$ is in cm$^{2}$/(V$\cdot$s) and $N_{i}$ is in /cm$^3$. So for ionized impurity scattering caused by heavy holes with $m_{h}^{\ast }$=0.28, we have
\begin{equation}
(\mu _{h})_{I}=\frac{1.08\times 10^{21}}{N_{i}}\left [ ln\frac{3.44\times 10^{18} }{N_{i}} \right ]^{-1},
\end{equation}  
and for the ionized impurity scattering caused by light holes with $m_{l}^{\ast }$=0.044, we have 

\begin{equation}
(\mu _{l})_{I}=\frac{2.72\times 10^{21}}{N_{i}}\left [ ln\frac{2.8\times 10^{20} }{N_{i}} \right ]^{-1},
\end{equation} 
so the total hole mobility caused by ionized impurity scattering can be calculated by Eq.\ref{e:com}:
\begin{equation}
\begin{split}
\mu_{I}&=((\mu _{l})_{I}+16.05(\mu _{h})_{I})/17.05\\
&=\frac{1.59\times 10^{20}}{N_{i}}\left [ ln\frac{2.8\times 10^{20} }{N_{i}} \right ]^{-1}+\frac{1.02\times 10^{21}}{N_{i}}\left [ ln\frac{3.44\times 10^{18} }{N_{i}} \right ]^{-1},
\end{split}
\label{e:mui2}
\end{equation} 

\begin{figure}
\centering
\includegraphics[angle=0,width=8.cm] {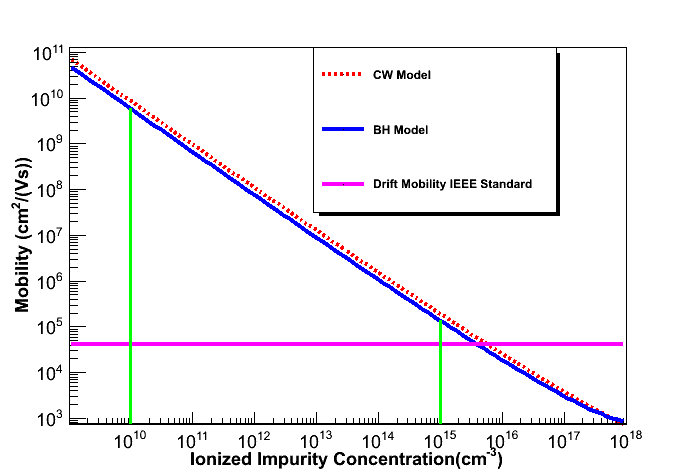}
\caption{The mobility contributed by ionized impurity scattering ($\mu_{I}$) as a function of the ionized impurity concentration. Note that when the ionized impurity concentration is in the region between 10$^{10}$/cm$^3$ and 10$^{15}$/cm$^3$, the total charge drift mobility contributed by the ionized impurity scattering is very small.   }
\label{fig:ionized}
\end{figure}

Eq.~\ref{e:mui2} indicates that $\mu_{I}$ decreases as $N_{i}$ increases at a given temperature. Fig.~\ref{fig:ionized} shows the relationship between the mobility due to ionized impurity scattering ($\mu_{I}$) and ionized impurity concentration.

The BH model has been well studied when $N_{i}$ is in the range of 10$^{14}$/cm$^3$-10$^{18}$/cm$^3$. Based on the IEEE Standard~\cite{IEEE}, the value of the charge drift mobility is $\mu_{n}$=36000 cm$^2$/(V$\cdot$s) and $\mu_{p}$=42000 cm$^2$/(V$\cdot$s) for n-type and p-type high-purity germanium crystals, respectively.
As shown in Fig.~\ref{fig:ionized}, it is clear that the total charge drift mobility is affected by the ionized impurities only when the ionized impurity concentration is higher than $\sim$10$^{14}$/cm$^{3}$. The ionized impurity concentration in detector-grade germanium crystal must be in the order of a few times 10$^{10}$/cm$^3$~\cite{EEHA}. With this very low ionized impurity concentration, the mobility due to ionized impurity scattering is in the order of 10$^{9}$ cm$^{2}$/(V$\cdot$s) as indicated in Fig.~\ref{fig:ionized}. Thus, the contribution from the ionized impurity scattering to the total charge drift mobility is very small in general.

\subsection{Acoustic phonon scattering}
At temperatures above absolute zero, the lattices of a germanium crystal are constantly vibrating, and the vibrations are treated as quasiparticles named phonons. In almost all semiconductors, there are two types of phonons: acoustic phonons (coherent movements of atoms of the lattice) and optical phonons (out-of-phase movements of the atoms in the lattice). The scattering of charge carriers by acoustical phonons is believed to be a very important contribution to the total charge drift mobility in semiconductors~\cite{dmbr}. The relaxation time for acoustic phonon scattering was originally derived by Bardeen and Shockley~\cite{jb} and can be written as:
\begin{equation}
\frac{1}{\tau }=\frac{\nu E_{ac}^{2}}{\pi \hbar^{4}c_{1}}m^{\ast 2}k_BT,
\label{e:mua}
\end{equation}
where $E_{ac}$=9.5 eV is the acoustic deformation potential constant and $c_{1}$=1.32$\times$ 10$^{12}$ dyn/cm$^{2}$ is the longitudinal elastic constant of germanium at 77 Kelvin along $<100>$ crystal orientation.
 
The mobility due to acoustic deformation potential scattering, $\mu_{A}$, can be calculated by the following equation~\cite{jb}:
\begin{equation}
{\mu}_{A}=\frac{2\sqrt{2\pi } \hbar^{4}c_{1}q}{3E_{ac}^{2}{m^{\ast 5/2}}(k_BT)^{3/2}},
\label{e:mua1}
\end{equation}
Again, with all constants in eq.~\ref{e:mua1} replaced by their values, eq.~\ref{e:mua1} becomes:
\begin{equation}
{\mu}_{A}=\frac{4.65\times 10^{5}}{m^{\ast 5/2}}\cdot T^{-3/2},
\end{equation}
where $\mu_{A}$ is in cm$^{2}$/(V$\cdot$s). With $m_{h}^{\ast }$=0.28 and $m_{l}^{\ast }$=0.044, we have 
\begin{equation}
(\mu _{h})_{A}=1.12\times 10^{7}\cdot T^{-3/2},
\end{equation}
and
\begin{equation}
(\mu _{l})_{A}=1.15\times 10^{9}\cdot T^{-3/2},
\end{equation}
so the total hole mobility caused by ionized impurity scattering obtained by Eq.\ref{e:com} is 
\begin{equation}
\mu_{A}=7.77\times 10^{7}\cdot T^{-3/2},
\label{e:mua2}
\end{equation}

\begin{figure}
\centering
\includegraphics[angle=0,width=8.cm] {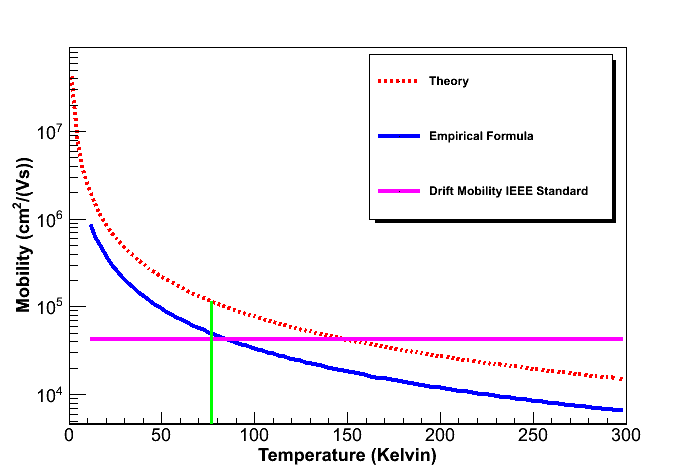}
\caption{The mobility due to acoustic phonon scattering ($\mu_{A}$) as a function of  temperature.}
\label{fig:acoustic}
\end{figure}

Eq.~\ref{e:mua2} implies that there is a temperature dependence in $\mu_{A}$. With $T$ = 77 Kelvin, one obtains $\mu_{A}$ = 1.15$\times$10$^{5}$cm$^{2}$/(V$\cdot$s). In D. M. Brown and R. Bray's work~\cite{dmbr}, the empirical expression for $\mu_{A}$ is:
\begin{equation}
{\mu}_{A}=3.37\times 10^{7}\cdot T^{-3/2},
\end{equation}
which differs about a factor of two from eq.~\ref{e:mua2}. Basing on their experimental work, they concluded that at low temperatures, acoustical phonon scattering is the main scattering mechanism in the presence of high electric fields. Their work didn't include the neutral impurity scattering in the analysis, only acoustical and ionized impurity scattering were taken into account. While treating acoustical phonon as sole adjustable parameter, they also pointed out that neutral impurity scattering might be important with high neutral impurity concentration. In this paper, we take into account acoustic phonon scattering, ionized impurity scattering  and the neutral impurity scattering to investigate their impacts on the charge drift mobility of high-purity germanium. Fig.~\ref{fig:acoustic} shows the variation of $\mu_{A}$ with temperature for both theoretical and empirical formulas. From Fig.~\ref{fig:acoustic}, it is clear that the theoretical results of $\mu_{A}$ are larger than the IEEE Standard at 77 Kelvin which indicates that acoustic phonon scattering may not be the sole scattering source of total charge drift mobility at 77K, there must be other scattering mechanisms contributing to the total charge drift mobility. The empirical formula fit the IEEE Standard well at 77 Kelvin if we take both neutral impurity scattering and acoustic phonon scattering into account.

\subsection{Neutral impurity scattering}
Neutral impurity scattering occurs when a charge carrier approaches close to a neutral impurity atom and exchanges its momentum with the atom. During the scattering, the charge carrier is deflected by the short-range interaction. In the literature, some authors consider that neutral impurity scattering may play a significant role in estimating the total charge drift mobility~\cite{co, nsp}. N. Sclar pointed out that neutral impurity scattering may be the dominant process or can be comparable with other scattering process in determining the total charge drift mobility, when the neutral impurity concentration is equal or greater than the ionized impurity concentration~\cite{nsp}. Therefore, in high-purity germanium crystals with ionized impurity concentration as low as 10$^{10}$/cm$^3$, there is a great possibility that neutral impurity scattering is an important scattering process in the determination of total charge drift mobility.

The neutral impurity scattering due to un-ionized donors or acceptors and neutral defects in a semiconductor were first derived by Erginsoy in 1950~\cite{ce}. He treated all neutral impurities as hydrogen atoms using the Bohr model. This approximation gave us the relaxation time for the neutral impurity scattering:
\begin{equation}
\frac{1}{\tau }=\frac{20\varepsilon_{r}\varepsilon_0 N_{n}h^{3}}{8\pi^{3}m^{\ast2}q^{2}},
\label{e:tau}
\end{equation}
where ${N}_{n}$ is the neutral impurity concentration. Combining eq.~\ref{e:tau} and eq.~\ref{mu}, we obtain the mobility in a given direction due to neutral impurity scattering $\mu_{1N}$:
\begin{equation}
\mu_{1N}= (\mu_{1N})_{E} =  \frac{q^{3}}{20N_{n}\hbar^{3}}\cdot \frac{m^{\ast }}{\varepsilon_{r}\varepsilon_0}.
\label{e:mun0}
\end{equation}
Since the neutral impurity scattering is an isotropic scattering process~\cite{tokm}, the charge carriers would be scattered with equal efficiency into all possible directions. Thus, eq.~\ref{e:mun0} can be modified as~\cite{rk}:
\begin{equation}
\mu_N = (\mu_{N})_{E} = \frac{q^{3}}{80\pi N_{n}\hbar^{3}}\cdot \frac{m^{\ast }}{\varepsilon_{r}\varepsilon_0}.
\label{e:mun2}
\end{equation}
The only difference between eq.~\ref{e:mun0} and eq.~\ref{e:mun2} is a factor of 4$\pi$ in the denominator of eq.~\ref{e:mun2} since the scattering process is isotropic~\cite{tokm}. Again, with all constants in eq.~\ref{e:mun2} replaced by their corresponding values, eq.~\ref{e:mun2} is reduced to be:
\begin{equation}
\mu_N = (\mu_{N})_{E} =\frac{8.95\times 10^{20}\cdot m^{\ast } }{N_{n}}.
\label{e:mun3}
\end{equation}
then considering heavy holes,
\begin{equation}
(\mu _{h})_{N}=\frac{2.51\times 10^{20} }{N_{n}}
\end{equation}
and light holes,
\begin{equation}
(\mu _{l})_{N}=\frac{3.94\times 10^{19} }{N_{n}}
\end{equation}
 which yields the total hole mobility caused by neutral impurity scattering is 

\begin{equation}
\mu_N=(\mu _{N})_E=\frac{2.31\times 10^{18}+2.36\times 10^{20}}{N_{n}}
\label{e:neu100}
\end{equation}

\begin{figure}
\centering
\includegraphics[angle=0,width=8.cm] {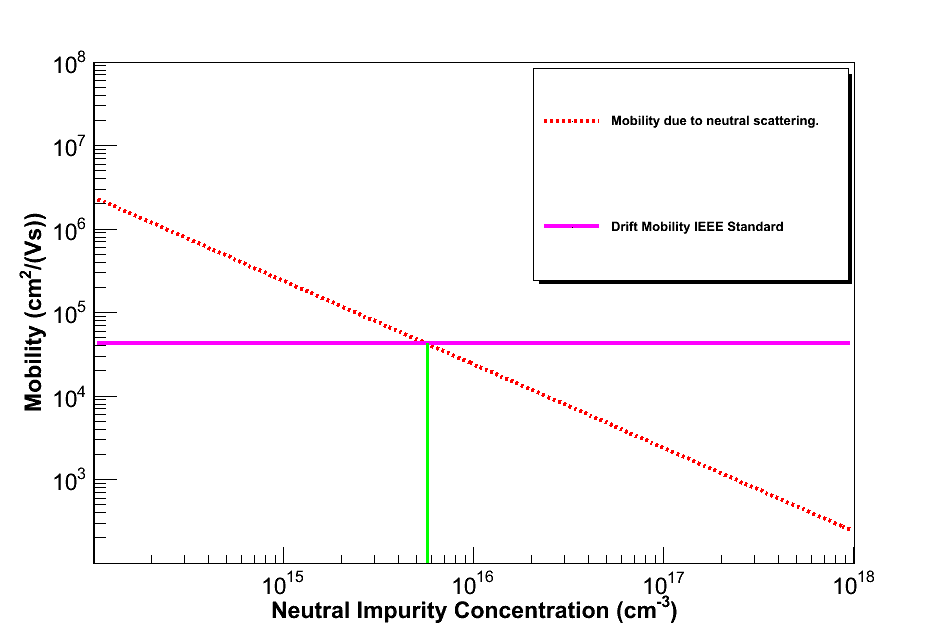}
\caption{The mobility due to neutral impurity scattering calculated by Erginsoy's model ($\mu_{N}$) as a function of  neutral impurity concentration. }
\label{fig:neu1}
\end{figure}

As shown in Fig.~\ref{fig:neu1}, the mobility due to neutral impurity scattering decreases as the neutral impurity concentration increases. With the neutral impurity concentration in the level between 10$^{14}$/cm$^3$-10$^{16}$/cm$^3$, the mobility due to neutral impurity scattering could be an important contribution to the total charge drift mobility at a level of close to the IEEE standard stated earlier.
From Erginsoy's work (eq.~\ref{e:neu100}), we can see that $\mu_{N}$ is temperature independent. Later on, Sclar~\cite{ns, tcm} gave another approximation for $\mu_{N}$, which showed that there is a weak dependence of $\mu_N$ on the temperature:
\begin{equation}
\mu_N=(\mu _{N})_S=0.82(\mu _{N})_E[\frac{2}{3}(\frac{k_{B}T}{E_{N}})^{1/2}+\frac{1}{3}(\frac{E_{N}}{k_{B}T})^{1/2}],
\label{e:neu111}
\end{equation}
where $(\mu _{N})_E$ is the temperature-independent mobility given by eq.~\ref{e:neu100} and $E_{N}$ is the scaled binding energy for the negative ion, $E_{N}$=0.71eV $m^{\ast }$/$m_{e}$($\varepsilon_{r}\varepsilon_0)^2$.

If we assume the neutral impurity concentration is 2$\times$10$^{15}$/cm$^3$ as measured by \cite{eeh}, then eq.~\ref{e:neu100} becomes a constant, 1.19$\times$10$^{5}$ cm$^{2}$/(V$\cdot$s), and eq.~\ref{e:neu111} becomes: 
\begin{equation}
\mu_N=(\mu _{N})_S=9.76\times 10^{4}(0.228T^{1/2}+0.976T^{-1/2}),
\label{e:neu2}
\end{equation}
which yields that $(\mu _{N})_S$= 2.06$\times$10$^{5}$ cm$^{2}$/(V$\cdot$s) at 77 Kelvin.

\begin{figure}
\centering
\includegraphics[angle=0,width=8.cm] {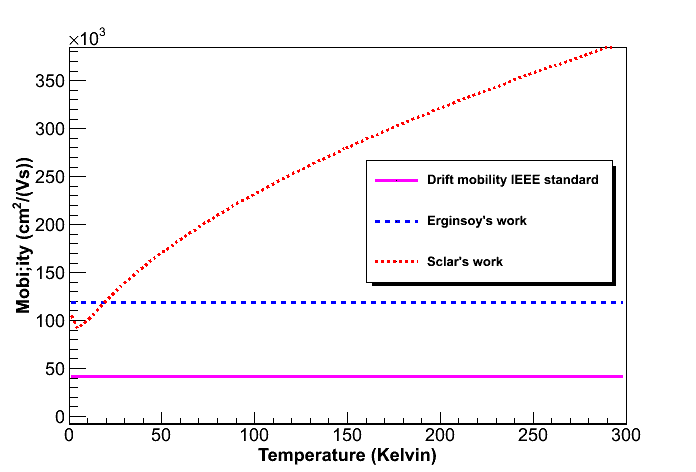}
\caption{The comparison among the work by Erginsoy, Sclar, and IEEE standard for mobility due to neutral impurity scattering with the assumption that the neutral impurity concentration is 2$\times$10$^{15}$/cm$^3$.}
\label{fig:neutral-com}
\end{figure}

\begin{figure}
\centering
\includegraphics[angle=0,width=8.cm] {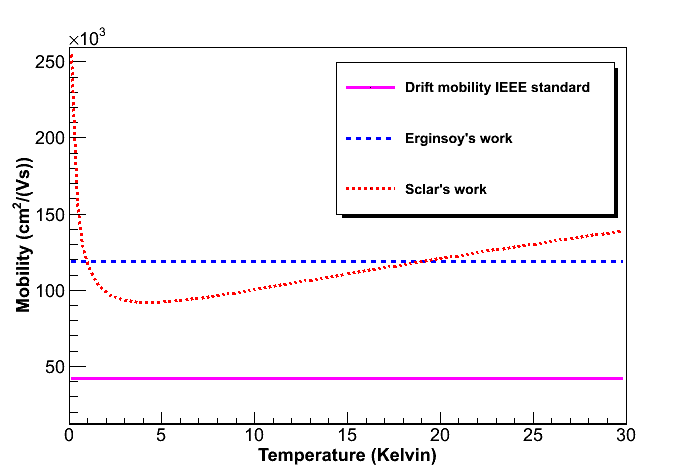}
\caption{The calculated freeze-out temperature of our germanium crystals is around 4 Kelvin.}
\label{fig:neutral-com.1}
\end{figure}

As shown in Fig.~\ref{fig:neutral-com}, there is a factor of two difference for the mobility due to neutral impurity scattering among the work by Erginsoy and Sclar at 77 Kelvin. At very low temperature, charge carrier freeze-out occurs in semiconductors and those shallow-level impurities become neutral and act like neutral impurity scattering centers. For germanium, the freeze-out temperature is below 20 Kelvin when the germanium crystal is pure enough~\cite{ppd}. Our calculation shows that the freeze-out temperature of our germanium crystals is around 4 Kelvin shown in Fig.~\ref{fig:neutral-com.1}. This means all of the neutral impurity scattering centers are from the original neutral impurities at 77 Kelvin. Thus, we followed Erginsoy's theory in this work, i.e. $\mu_{N}$ has no temperature dependence.

\subsection{Other scatterings}
For the mobility caused by the scattering of optical phonons in germanium, $\mu_{O}$, it is negligible for two reasons. Firstly, $\mu_{O}$ is temperature dependent and it only becomes important for silicon and germanium above room temperature, 300 Kelvin~\cite{bs}. Therefore, for germanium detectors at 77 Kelvin, $\mu_{O}$ has almost no contribution to the total charge drift mobility and thus can be ignored. Secondly, optical phonon scattering is generally not important for charge carriers in the conduction band along $<100>$ germanium crystal according to the work in~\cite{ssl}. Thus, $\mu_{O}$ can be ignored for our high-purity germanium crystal along the $<100>$ orientation. 

For dislocation scattering (also known as defects scattering) in germanium, it is generally known that this has big impact on the total charge drift mobility only when the dislocation density is greater than the order of 10$^{7}$/cm$^2$~\cite{dt}. However, the dislocation density of germanium crystals for radiation detectors is usually on the level of 300-10,000 /cm$^2$~\cite{gwy}. With such low dislocation density, dislocations cannot be important scattering centers in comparison with other scattering mechanisms mentioned above. This means the contribution from $\mu_{D}$ to the total mobility ($\mu_{T}$) can be ignored as well.

\section{Experimental Results}
\label{s:exp}
High-purity germanium crystals have been grown on a weekly basis in our labs at the University of South Dakota~\cite{guo1, gang1, gang2, guo3, gang3, gang4, gang5, guo4}. Several germanium samples obtained from a detector-grade crystal (NO.20~\cite{gj}) grown in our lab with measured Hall mobility $\mu_{H}$ larger than 36000 cm$^2$/(V$\cdot$s) were used for our investigation in this work. The relationship between the Hall mobility $\mu_{H}$ and the total charge drift mobility $\mu$ is defined as~\cite{IEEE}:
\begin{equation}
\mu=\frac{\mu _{H}}{r},
\label{e:mun4}
\end{equation}
where $r$ is a constant near unity. Based on IEEE Standard~\cite{IEEE}, $r$ is assumed to be 0.83 and 1.03 for n-type and p-type high-purity germanium crystals, respectively. 

\begin{figure}
\centering
\includegraphics[angle=0,width=8.cm] {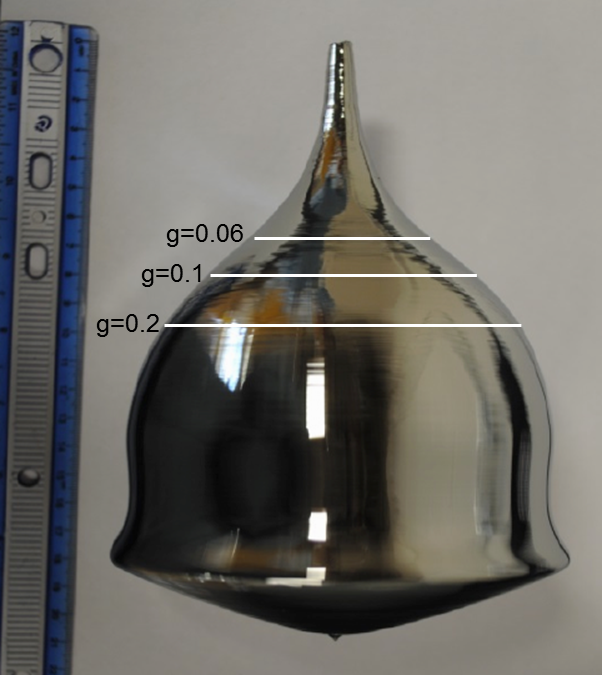}
\caption{Three wafers were cut from the detector grade crystal NO.20 with the white lines indicating the position. The parameter, $g$,  denotes the fraction of original liquid which is frozen. }
\label{fig:cut1}
\end{figure}

\begin{figure}
\centering
\includegraphics[angle=0,width=8.cm] {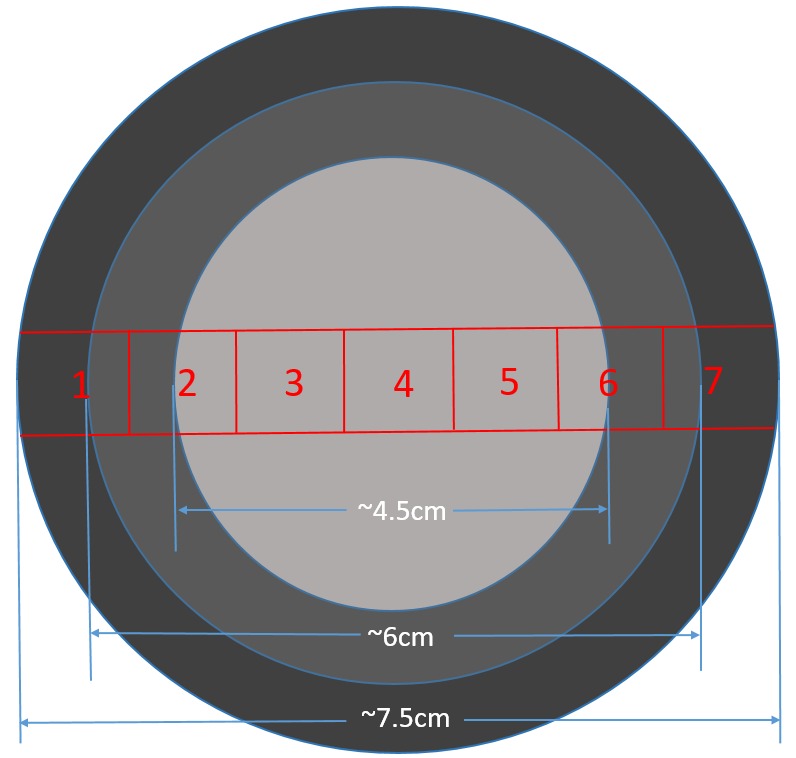}
\caption{Schematic diagram showing the location of the samples cut from the three wafers.}
\label{fig:cut2}
\end{figure}

Three wafers were cut from the detector-grade crystal mentioned above at different axial positions, denoted by g, the fraction of the melt that has crystallized, as shown in Fig.~\ref{fig:cut1}. Five square samples were cut from the wafer with g=0.06, seven samples were cut from the wafer with g=0.1 and another seven samples were cut from the wafer with g=0.2 as shown in Fig.~\ref{fig:cut2}. These samples with area of $\sim$1 cm$^{2}$ and thickness of 1.5mm were etched with etchant (HF: HNO$_{3}$=1:3), rinsed with deionized water and blew dried with nitrogen gas. Then four In-Ga eutectic ohmic contacts are scratched onto the four corners of the samples before we measured their electrical properties, such as ionized impurity concentration, Hall mobility and resistivity, by using the Van der Pauw Hall Effect Measurement System at 77 Kelvin. For uniform samples, the uncertainty is determined to be less than 5\% in the resistivity and Hall coefficient measurement~\cite{kab}. In this work, 5\% was applied to all data points as the uncertainty for the Hall effect measurements.

\begin{figure}
\centering
\includegraphics[angle=0,width=8.cm] {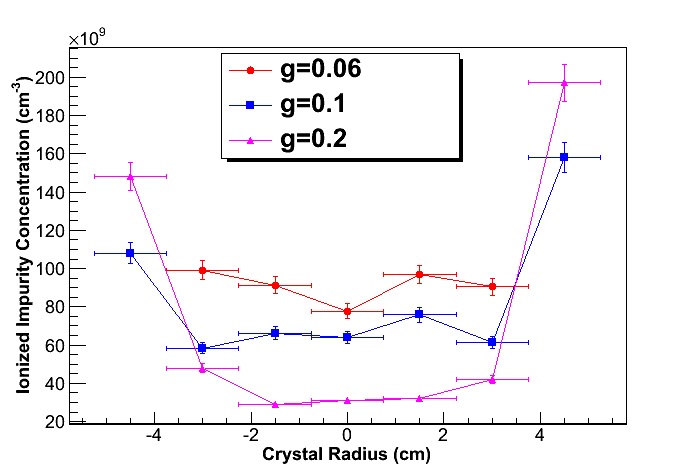}
\caption{The net carrier concentration as a function of crystal radius for all germanium samples.}
\label{fig:concentration}
\end{figure}

Fig.~\ref{fig:concentration} shows the ionized impurity concentration as a function of crystal radius for all germanium samples. 
As indicated by Fig.~\ref{fig:concentration}, the closer the sample to the center of the crystal, the lower ionized impurity concentration it has. The crystal in the central part has a better impurity level than the edge part.

\begin{figure}
\centering
\includegraphics[angle=0,width=8.cm] {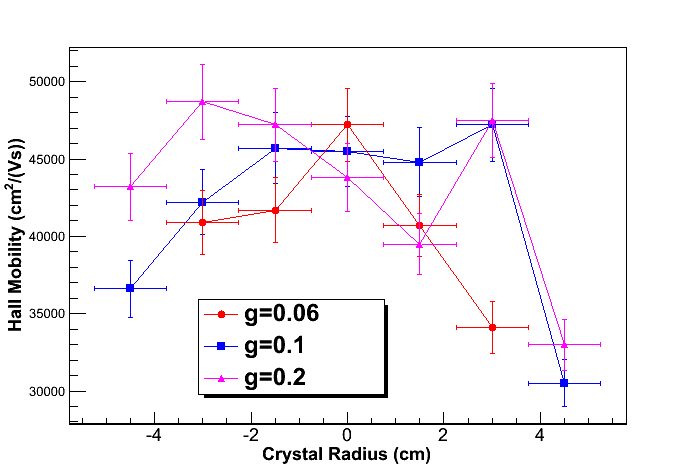}
\caption{The radial distribution of Hall mobility of the three wafers.}
\label{fig:mobility}
\end{figure}

\begin{figure}
\centering
\includegraphics[angle=0,width=8.cm] {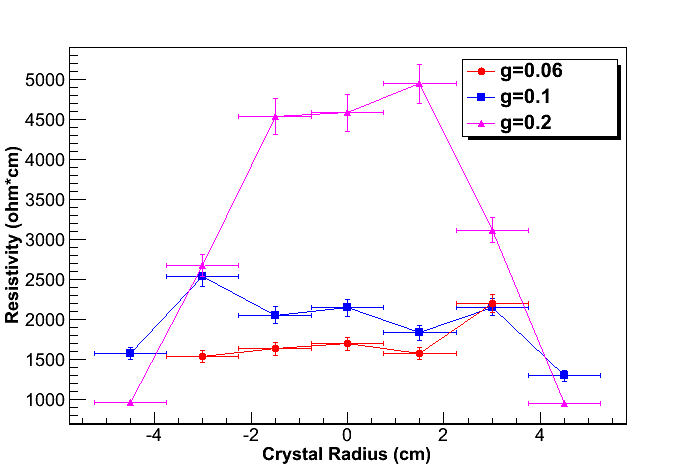}
\caption{The radial distribution of resistivity of the three wafers.}
\label{fig:resistivity}
\end{figure}

Fig.~\ref{fig:mobility} and Fig.~\ref{fig:resistivity} show the radial distribution of Hall mobility and resistivity for all samples, respectively. From Fig.~\ref{fig:mobility} and Fig.~\ref{fig:resistivity}, we can see that the mobility and resistivity at the edge of the crystal are smaller than that at the central part of the crystal.

According to our Hall Effect measurements shown in Fig.~\ref{fig:mobility}, our Ge crystals have a total charge drift mobility $\mu_{T}$ of $\sim$ 42000 cm$^2$/(V$\cdot$s) with an ionized impurity concentration in the range of 10$^{10}$/cm$^3$-10$^{11}$/cm$^3$ as shown in Fig.~\ref{fig:concentration}. Using eq.~\ref{e:mui2}, the calculated mobility due to this level of the ionized impurity scattering $\mu_{I}$ is in the range of 6.61$\times$10$^{8}$-5.85$\times$10$^{9}$ cm$^{2}$/(V$\cdot$s). Similarly, from eq.~\ref{e:mua2}, when $T$ = 77 Kelvin, the mobility due to acoustic phonon scattering $\mu_{A}$ = 1.15$\times$10$^{5}$cm$^{2}$/(V$\cdot$s). With $\mu_{O}$ as well as $\mu_{D}$ being ignored, the mobility due to neutral impurity scattering $\mu_{N}$ can be deduced using eq.~\ref{e:mu}. Our deduced results showed that $\mu_{N}$ is almost a constant, 6.6$\times$ 10$^{4}$cm$^{2}$/(V$\cdot$s).  Since the deduced $\mu_{N}$ and the calculated $\mu_{A}$ are much smaller than $\mu_{I}$ for a detector-grade crystal at 77 Kelvin, we conclude that the total charge drift mobility ($\mu_{T}$) in our germanium crystals is dominated by both $\mu_{N}$ and $\mu_{A}$, and the neutral impurity concentration has important impact on the charge drift mobility.

\begin{figure}[htb!]
\centering
\includegraphics[angle=0,width=8.cm] {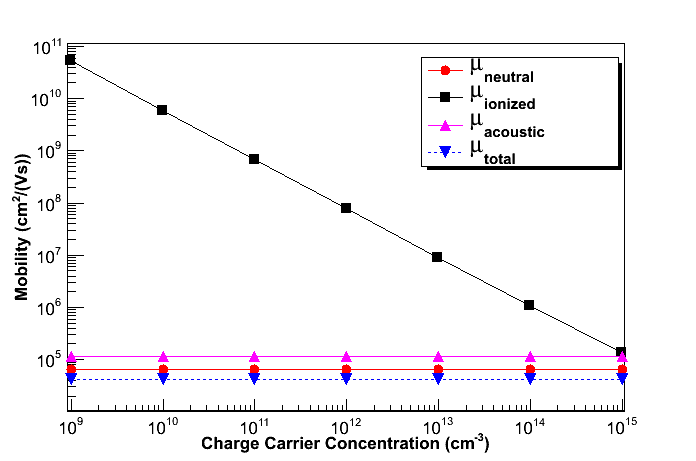}
\caption{The relationship between the charge carrier concentration and all the scattering processes at 77 Kelvin.}
\label{fig:den}
\end{figure}

\begin{figure}[htb!]
\centering
\includegraphics[angle=0,width=8.cm] {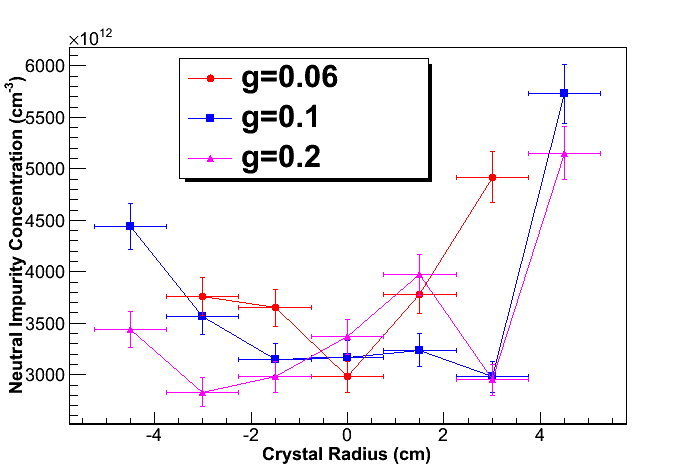}
\caption{The radial distribution of neutral impurity concentration.}
\label{fig:net}
\end{figure}

Once $\mu_{N}$ is obtained, the neutral impurity concentration, $N_n$, can then be calculated from eq.~\ref{e:neu100}, which yields that $N_n$ is in the range of 2.8$\times$10$^{15}$/cm$^3$-5$\times$10$^{15}$/cm$^3$ when the ionized impurity concentration is in the range of 10$^{10}$/cm$^3$-10$^{11}$/cm$^3$. This level of neutral impurity concentration agrees with
the pioneering work (2$\times$10$^{15}$/cm$^{3}$) by W. L. Hansen, E. E. Haller, and P. N. Luke in 1982~\cite{wlh}.
Fig.~\ref{fig:net} shows the relationship between the calculated neutral impurity concentration ($N_n$) and the crystal radius. Fig.~\ref{fig:net} implies that there are more neutral impurities at the edge than in the center of the crystal. This is very similar to the case of the radial distribution of ionized impurities, where the neutral impurity is different, the total charge drift mobility can be different. 

In D. M. Brown and R. Bray's work~\cite{dmbr}, they also summarized the properties of germanium samples at both 77 Kelvin and 300 Kelvin with carrier concentration larger than 10$^{13}$/cm$^3$. We assume  our equations about $\mu_{I}$ and $\mu_{A}$ are correct, then we use their measured ionized impurity concentration to calculate $\mu_{I}$, then $\mu_{N}$ can be determined with the known $\mu_{I}$, $\mu_{A}$ and $\mu_{T}$ from their data. Neutral impurity concentration $N_{n}$ can be calculated by eq.~\ref{e:mun3} to see a variation of neutral impurity shown in table~\ref{tab:table2}. The obtained neutral impurity concentration from these measurements are from 3.7$\times$10$^{15}$ to 7.77$\times$10$^{15}$/cm$^3$, which is very close to the calculated neutral impurity concentration in our crystals. 

\begin{table*}[t]
\centering
\caption{Calculated $\mu_{N}$ and neutral impurity concentration $N_{n}$, using D. M. Brown and R. Bray's data}
\label{tab:table2}
\begin{tabular}{|c|c|c|c|}
\hline
 $N_{A}-N_{D}$ (/cm$^{3}$) & $\mu_{measured}$ (cm$^2$/(V$\cdot$s)) & $\mu_{N}$ (cm$^2$/(V$\cdot$s)) & $N_{n}$ (/cm$^3$)\\
\hline 
 1.25$\times$10$^{13}$ & (4.12$\pm$0.2)$\times$10$^{4}$ & (6.48$\pm$0.32)$\times$10$^{4}$ & (3.68$\pm$0.06)$\times$10$^{15}$           \\
\hline
  1.52$\times$10$^{14}$ & (3.03$\pm$0.15)$\times$10$^{4}$ & (4.36$\pm$0.22)$\times$10$^{4}$ & (5.47$\pm$0.13)$\times$10$^{15}$           \\
\hline
  1.81$\times$10$^{14}$ & (3.40$\pm$0.17)$\times$10$^{4}$ & (5.23$\pm$0.49)$\times$10$^{4}$ & (4.56$\pm$0.09)$\times$10$^{15}$           \\
\hline
 2.65$\times$10$^{14}$ & (2.99$\pm$0.15)$\times$10$^{4}$ & (4.44$\pm$0.36)$\times$10$^{4}$ & (5.37$\pm$0.13)$\times$10$^{15}$           \\
 \hline
 9.75$\times$10$^{14}$ & (2.26$\pm$0.11)$\times$10$^{4}$ & (3.51$\pm$0.23)$\times$10$^{4}$ & (6.78$\pm$0.20)$\times$10$^{15}$           \\
 \hline
 1.51$\times$10$^{15}$ & (1.95$\pm$0.10)$\times$10$^{4}$ & (3.11$\pm$0.19)$\times$10$^{4}$ & (7.67$\pm$0.25)$\times$10$^{15}$          \\
 \hline
 2.23$\times$10$^{15}$ & (1.83$\pm$0.09)$\times$10$^{4}$ & (3.19$\pm$0.18)$\times$10$^{4}$ & (7.46$\pm$0.26)$\times$10$^{15}$          \\
\hline
\end{tabular}
\end{table*}

There are many sources for the neutral impurities during germanium crystal purification and growth processes. Carbon as the neutral impurity may be introduced into the purified ingot from both the graphite boat and the coating layer of the inside surface of the quartz boat during zone refining process with concentration about 10$^{13}$/cm$^3$~\cite{eeh}. Additionally, during crystal growth, high-purity germanium crystals are grown in hydrogen atmosphere~\cite{gwy}. The solubility of hydrogen in germanium at its melting temperature is 4$\times$ 10$^{14}$/cm$^3$~\cite{wlh}. The quartz crucible is used to hold the germanium melt. Silica can react with hydrogen and germanium melt to form silicon and oxygen at the germanium melting temperature. The concentrations of silicon and oxygen in high-purity germanium crystal are at similar level of 10$^{14}$/cm$^3$~\cite{rjf} which is consistent with our calculation.

\section{Conclusions}
\label{s:conc}
We investigated the scattering mechanisms that contribute to the total charge drift mobility in detector-grade germanium crystals. Using the measured Hall mobility and the calculated mobility due to ionized impurity and acoustic phonon scatterings, we deduced the mobility contributed by the neutral impurity scattering using the Matthiessen's rule. As a result, the neutral impurity concentration is evaluated. Based on our calculation, we found that for high-purity germanium crystal along $<100>$ direction, with impurity level of 10$^{10}$/cm$^3$-10$^{11}$/cm$^3$ and dislocation density below 10$^{4}$/cm$^2$, the neural impurity scattering is an important scattering mechanism at 77 Kelvin. The neutral impurity concentration has a large impact on the charge drift mobility. There are more neutral and ionized impurity atoms at the edge parts of the crystal than that at the center part, which results in the lower charge drift mobility at the edge part and higher charge drift mobility at the center part. 

\section*{Acknowledgments}
The authors wish to thank Christina Keller and David Radford for their careful reading of this manuscript. We also would like to thank Jing Liu and Arun Soma for their useful discussion.  This work was supported in part by NSF PHY-0919278, NSF PHY-1242640, NSF OIA 1434142, DOE grant DE-FG02-10ER46709, the Office of Research at the University of South Dakota and a research center supported by the State of South Dakota.

\end{document}